\documentclass[aps,prb,floatfix,amsmath,amssymb,eqsecnum,nofootinbib,twocolumn,superscriptaddress]{revtex4}

\usepackage{amsmath,amssymb,dsfont,graphicx,psfrag}
\usepackage{epsfig}
\usepackage{graphicx}
\usepackage{epstopdf}

\def\beq{\begin{equation}}
\def\eeq{\end{equation}}

\begin{document}
\title{Quantum dynamics of a domain wall in a quasi one-dimensional $XXZ$ ferromagnet}

\author{Pavel Tikhonov}
\affiliation{Department of Physics, Bar-Ilan University, Ramat-Gan 52900, Israel}

\author{Efrat Shimshoni}
\affiliation{Department of Physics, Bar-Ilan University, Ramat-Gan 52900, Israel}

\begin{abstract}
We derive an effective low-energy theory for a ferromagnetic $(2N+1)$-leg spin-$\frac{1}{2}$ ladder with strong $XXZ$ anisotropy $\left|J_{\parallel}^z\right|\ll \left|J_{\parallel}^{xy}\right|$, subject to a kink-like non-uniform magnetic field $B_z(X)$ which induces a domain wall (DW). Using Bosonization of the quantum spin operators, we show that the quantum dynamics is dominated by a single one-dimensional mode, and is described by a sine-Gordon model. The parameters of the effective model are explored as functions of $N$, the easy-plane anisotropy $\Delta=-J_{\parallel}^z/J_{\parallel}^{xy}$, and the strength and profile of the transverse field $B_z(X)$. We find that at sufficiently strong and asymmetric field, this mode may exhibit a quantum phase transition from a Luttinger liquid to a spin-density-wave (SDW) ordered phase. As the effective Luttinger parameter grows with the number of legs in the ladder ($N$), the SDW phase progressively shrinks in size, recovering the gapless dynamics expected in the two-dimensional limit $N\rightarrow\infty$.

\end{abstract}
\pacs{75.10.Pq,75.10.Jm,75.30.Kz}
\maketitle

\section{Introduction}
Low dimensional quantum systems attract much experimental and theoretical attention, due to the rich physics arising from their enhanced quantum fluctuations. Most prominently, quantum effects are manifested by quasi-one dimensional (1D) spin systems, namely, spin ladders \cite{LadderReview}. Theoretical studies primarily focused on models of coupled spin-$\frac{1}{2}$ chains with anti-ferromagnetic (AFM) exchange interactions. In addition to being motivated by the existence of real materials, these were inspired by the seminal work of Haldane \cite{HaldaneConjecture}, which pointed out the crucial distinction between odd and even spin $S$. The same physics carries through to multi-leg spin-$\frac{1}{2}$ ladder systems \cite{Schultz,Affleck,Shelton}, where the number of legs $N$ replaces the general spin $S=\frac{N}{2}$ in the Haldane chain.
A particularly striking signature of quantum fluctuations arises in the case of odd $N$, characterized by a magnetically disordered ground-state with power-law spin-spin correlations. This behavior, which reflects the presence of gapless spin-flip excitations (spinons), is the simplest realization of a so-called "spin-liquid" phase \cite{balents-2010} beyond 1D.

Notably, the above described quantum features are characteristic to AFM spin ladders. Ferromagnetic (FM) ladders, on the other hand, are more "classical" in nature and in the Heisenberg ($SU(2)$-symmetric) case, long range FM order is established. Nevertheless, quantum effects were observed in certain spin-ladder compounds \cite{AFMMaterials}, and were shown theoretically to yield a rich phase diagram \cite{FMLadder1,FMLadder2}. A key ingredient promoting quantum fluctuations in such systems is anisotropy in the exchange interactions of the XXZ type. As a result, the spin system becomes either an easy-plane or easy axis FM, and can undergo a quantum phase transition, e.g. in the presence of a transverse field.

Realistic magnetic materials quite often possess Heisenberg exchange interactions, hence much of the earlier literature on quantum magnetism did not regard anisotropy as a significant parameter. However, the coupling of real spin to additional degrees of freedom may introduce appreciable (and possibly tunable) anisotropic interactions. In particular, in the recent years there has been growing interest in systems based on heavy elements with strong spin-orbit coupling, which pave the way to realizing a variety of unconventional spin models. A prominent example is the proposal \cite{SOplusCoulomb} to realize spin interaction terms in Iridate crystals that are effectively consistent with the highly anisotropic Kitaev model \cite{KitaevModel}, and hence support a quantum spin-liquid ground state in a genuinely two-dimensional (2D) system.

An alternative route to the formation of anisotropic exchange interactions naturally arises in systems with iso-spin degrees of freedom, such as the layer or valley index in bi-layers or bipartite lattices. A fascinating playground for such realizations of quantum spin models is provided by quantum Hall ferromagnetism (QHFM)\cite{QHFM}, established in 2D electron systems subject to a strong magnetic field. Most notable is its manifestation in graphene: the multi-component nature of the spin/iso-spin manifold leads to a plethora of exchange-induced broken symmetry phases, where anisotropy plays a crucial role \cite{Kharitonov_bulk}. Yet another type of system effectively described by inherently anisotropic quantum spin models is the superconducting (SC) ladder \cite{SupercondChains}, which implements a mapping of the complex SC order parameter field to a Bosonic representation of local spin operators.

A particularly appealing aspect of the latter two realizations is that the control of parameters, as well as the measurement of physical properties, are accessible by electric means. Specifically in graphene QHFM, electric conduction distinguishes a FM order in the bulk from other broken-symmetry phases \cite{Maher2013,Young2013} as it supports a gapless conducting mode. This mode is associated with quantum fluctuations of a domain wall (DW) configuration \cite{ShibataTakagi}, which forms at the edge of the sample in the FM state and can be modeled as a Luttinger liquid (LL) \cite{FertigBrey}. Due to the spin-charge coupling characteristic to quantum Hall systems, this mode is relatively protected from backscattering and exhibits a nearly perfect electrical conductivity \cite{us1416}. Similar conducting DW channels can also form in the bulk, e.g. in bilayer graphene where a non-uniform effective Zeeman field is induced by a spatially dependent gate voltage \cite{Mazo14,Kusum}. Derivation of an effective low energy theory describing the dynamics of these modes (and hence their transport properties) requires the quantum analysis of a spin system subject to non-uniform fields \cite{us1416,Kharitonov_new}.

In this paper we address the problem of a generic spin configuration in the quantum regime, where spin fluctuations can not be treated in the framework of spin wave approach \cite{SpinBooks}. To this end, we analyze the low energy dynamics of a smooth DW of an arbitrary finite width and shape, in a FM with anisotropic exchange interactions. This generalizes an earlier study of a sharp DW configuration \cite{MazoFertigShimshoni}, which has been predicted to possess a phase transition to a spin-ordered configuration as a function of anisotropy strength. In the present work we model the DW as a multi-leg ladder of ferromagnetic $XXZ$ spin-$\frac{1}{2}$ chains subject to a kink-like magnetic field $B_z(X)$, and derive a 1D low-energy effective theory which allows to explore the dependence of its quantum dynamics on field strength and shape. In particular, we identify the regime of parameters where a quantum phase transition from a LL to spin density wave (SDW) can occur.

The paper is organized as follows: In Sec. \ref{sec:Model} we present a model of coupled ferromagnetic spin chains and identify the $U(1)$ mode that dominates the low-energy behavior. In Sec. \ref{sec:EffectiveTheory} we detail the derivation of its effective low energy theory, and the implied phase diagram. Finally, we present concluding remarks in Sec. \ref{sec:sum_remarks}.

\section{The Model}
\label{sec:Model}
We consider a quasi-$1D$ model for a FM stripe as   a $(2N+1)$-leg ladder of coupled XXZ spin-$\frac{1}{2}$ chains in a non-uniform magnetic field oriented along the $z$-axis, that exhibits a sign reversal (see Fig. 1).
\begin{figure}
\label{fig:ladders}
\includegraphics[scale=0.6]{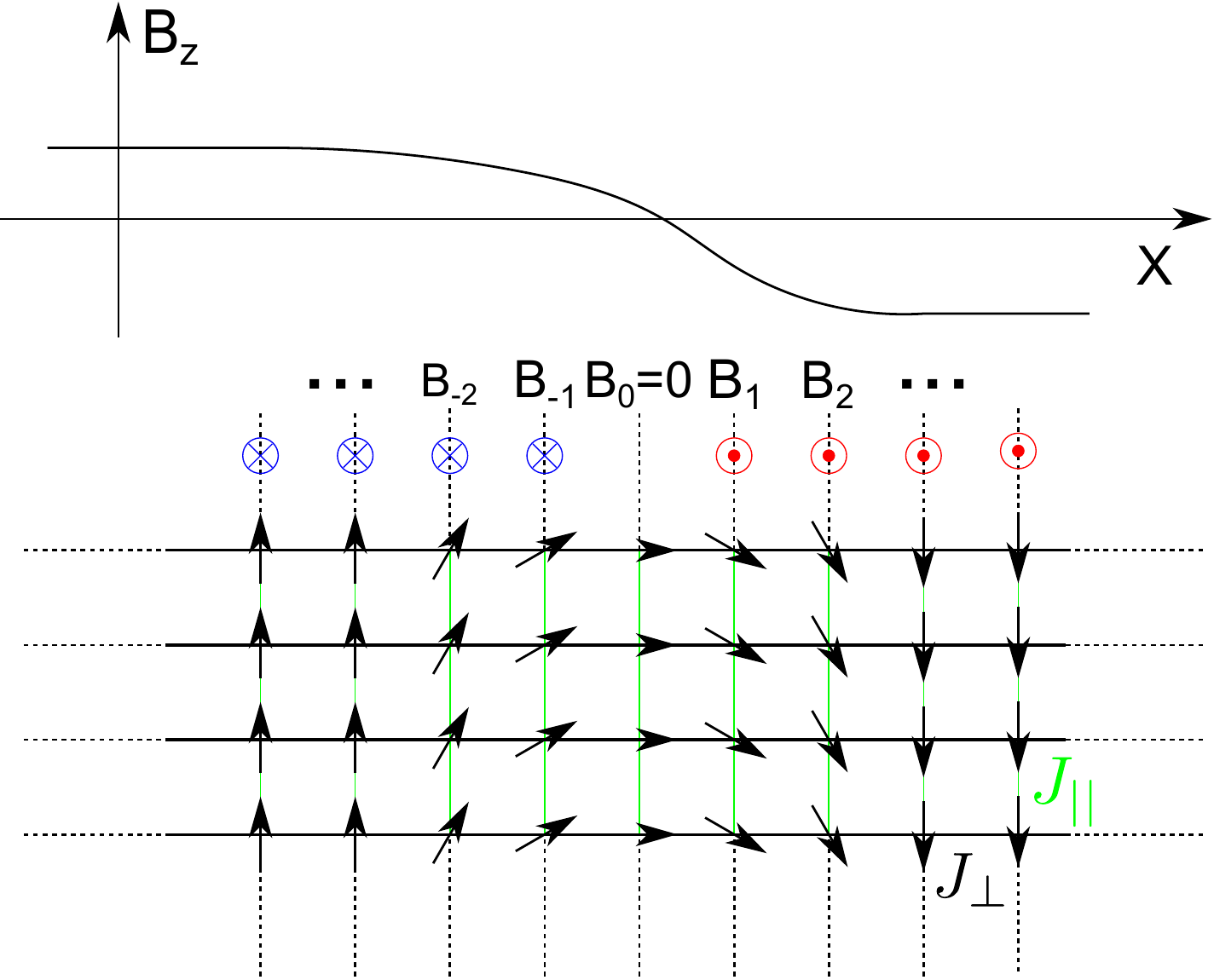}
\caption{(Color online.) Schematic representation of a domain wall configuration in a system described by Eq. (\ref{eq:H_original}); here $X$ is a continuum representation of the chain index $i$.}
\end{figure}
The Hamiltonian describing this system is
\begin{align}
\label{eq:H_original}
H	&=\sum_{i}H_{i}+H^{int}_{i}\\
H_{i}	&=\sum_{j}\frac{J_{||}^{xy}}{2}\left(S_{j,i}^{+}S_{j+1,i}^{-}+h.c.\right)+J_{||}^{z}S_{j,i}^{z}S_{j+1,i}^{z}-B_{i}S_{j}^{z},\nonumber \\
H^{int}_{i}	&=\sum_{j}\frac{J_{\perp}^{xy}}{2}\left(S_{j,i}^{+}S_{j,i+1}^{-}+h.c.\right)+J_{\perp}^{z}S_{j,i}^{z}S_{j,i+1}^{z},\nonumber \\
\nonumber
\end{align}
where all the coupling constants are ferromagnetic ($J_{\parallel}^z,J_{\perp}^z,J_{\perp}^{xy}<0$, while for convenience $J_{\parallel}^{xy}>0$  [\onlinecite{SpinRot}]), $i\in -N,\dots,-1,0,1,\dots,N$ is the leg index, $j$ is a site index along the chain and $B_{i}$ is a magnetic field acting on chain $i$. The natural degrees of freedom of unpolarized spins are Euler angles, which can be easily introduced by bosonization \cite{Giamarchi}. Taking the continuum limit along the chains ($S^{\nu}_{j,i}\rightarrow S^{\nu}_i \left(x\right)$),
\begin{align}
\label{eq:Spin_operators}
S_{i}^{\pm}\left(x\right)&=\frac{e^{\mp i\theta_{i}\left(x\right)}}{\sqrt{2\pi\alpha}}\left[\cos\left(2\varphi_{i}\left(x\right)\right)+\cos\left(2k_{F}x\right)\right]\\
S_{i}^{z}\left(x\right)&=-\frac{1}{\pi}\partial_x\varphi_{i}\left(x\right)+\frac{1}{\pi\alpha}\cos\left(2\varphi_{i}\left(x\right)-2k_{F}x\right)
\end{align}
where $\left[\varphi_i\left(x\right),\frac{1}{\pi}\partial_{x}\theta_j\left(x'\right)\right]=i\delta\left(x-x'\right)\delta_{ij}$, $\alpha$ is a short distance cut-off (lattice spacing) and $k_F=\frac{\pi}{2\alpha}$.
In terms of the new fields, the original Hamiltonian has the following form:
\begin{widetext}
\begin{align}
\label{eq:H_bosonized_long}
H&=\sum_{i=-N}^{N}H_{i}+\sum^{\pm N}_{i=\pm 1\dots}H^{int}_{i}, \\
H_{i}&=\frac{1}{2\pi}\int\mathrm{d}x\left[uK\left(\partial_{x}\theta_{i}\right)^{2}+\frac{u}{K}\left(\partial_{x}\varphi_{i}\right)^{2}\right]+\frac{2g_{3}}{\left(2\pi\alpha\right)^{2}}\int\mathrm{d}x\cos\left(4\varphi_{i}-4k_{F}x\right)+\frac{1}{\pi}B_{i}\partial_{x}\varphi_{i}-\frac{B_{i}}{\pi\alpha}\cos\left(2\varphi_{i}-2k_{F}x\right) \nonumber \\
H^{int}_{\pm i}&=\int\mathrm{d}x\Bigg[\frac{2g_{1}}{\left(2\pi\alpha\right)^{2}}\cos\left(\theta_{\pm i}-\theta_{\pm\left(i-1\right)}\right)+\frac{J_{\perp}^{z}\alpha}{\pi^{2}}\partial_{x}\varphi_{\pm i}\partial_{x}\varphi_{\pm\left(i-1\right)} \nonumber\\
&+\frac{2g_{2}}{\left(2\pi\alpha\right)^{2}}\cos\left(2\varphi_{\pm i}-2\varphi_{\pm\left(i-1\right)}\right)+\frac{2g_{2}}{\left(2\pi\alpha\right)^{2}}\cos\left(2\varphi_{\pm i}+2\varphi_{\pm\left(i-1\right)}-4k_{F}x\right)\Bigg] \nonumber,
\end{align}
\end{widetext}
and  $g_{1}=\pi J_{\perp}^{xy}\alpha$, $g_{2}=J_{\perp}^{z}\alpha$, $g_{3}=J_{\parallel}^{z}\alpha$; The Luttinger liquid (LL) parameters $K$ and $u$ are obtained exactly for an XXZ-chain with $\left|J_{\parallel}^{z}\right|< J_{\parallel}^{xy}$ by a Bethe-ansatz calculation \cite{LutherPeshelHaldane}, yielding for zero magnetic field
\begin{align}
K&=\frac{\pi}{2\arccos\left(\Delta\right)}, \\
u&=\frac{1}{1-\frac{1}{2K}}\sin\left(\pi\left(1-\frac{1}{2K}\right)\right)\frac{J_{\parallel}^{xy}}{2} \nonumber \\
\Delta&\equiv-\frac{J_{\parallel}^{z}}{J_{\parallel}^{xy}}, \nonumber
\end{align}
so that for ferromagnetic interactions $K>1$. It is important to point out that this result remains a good approximation even in a finite magnetic field for $\left|J_{\parallel}^z \right| \ll J_{\parallel}^{xy}$. The Gaussian part of the Hamiltonian (\ref{eq:H_bosonized_long}) can be conveniently written in a matrix form
\begin{align}
\label{eq:H_Gauss}
H_{Gauss}&=\frac{1}{2\pi}\int\mathrm{d}x\left[uK\partial_{x}\theta_{i}\delta_{ij}\partial_{x}\theta_{j}+\frac{u}{K}\partial_{x}\varphi_{i}\delta_{ij}\partial_{x}\varphi_{j}\right] \nonumber \\
&+\frac{J_{\perp}^{z}\alpha}{2\pi^{2}}\int\mathrm{d}x\partial_{x}\varphi_{i}C_{ij}\partial_{x}\varphi_{j},
\end{align}
where the matrix $C$ couples adjacent modes and is defined  as
\begin{align}
C_{ij}\equiv\delta_{i,j-1}+\delta_{i-1,j}
\end{align}
in which Einsteinian summation over $i,j$ is implied.

Being in the ferromagnetic region $\left( K>1 \right)$, the terms of the form
$\cos\left(\theta_{\pm\mu}-\theta_{\pm\left(\mu-1\right)}\right)$ (parametrized by the coupling constant $g_1$) are the most relevant among the non-Gaussian contributions to $H$, namely, with the lowest scaling dimension:
\begin{align}
d\left(\cos\beta\theta_{i}\right)\simeq\frac{\beta^2}{4K}
\end{align}
and hence
\begin{align}
d_{g_1}=\frac{1}{2K}<\frac{1}{2}.
\end{align}
Therefore we include these leading operators in the low-energy theory and account for all other non-quadratic terms in Eq. (\ref{eq:H_bosonized_long}) as perturbations. We then perform a transformation $D$ which simplifies the unperturbed Hamiltonian. The matrix $D$ can be related to a canonical transformation $U$ on the $K$-matrix of coupled Luttinger liquids. Here we keep the $\varphi$ and $\theta$ fields separate as in the original Hamiltonian (\ref{eq:H_bosonized_long}), and define
\begin{align}
\label{eq:transform}
&\bar{\varphi}_{i} =\left(D^{T}\right)^{-1} _{ij}\varphi_{j}, \quad \bar{\theta}_{i} =D_{ij}\theta_{j}, \nonumber \\
&D_{n,m}	=\delta_{n,m}+\delta_{n,0}-\delta_{n-sg\left(n\right),m},
\end{align}
where $D$ is the $\left(2N+1\right)\times\left(2N+1\right)$ matrix, $n$ is a row number, $m$ is a column number, $-N<n,m<N$ and
\begin{align}
sg\left(m\right)=\begin{cases}
1 & m>0\\
0 & m=0\\
-1 & m<0
\end{cases}.
\end{align}
Note that the transformation Eq. (\ref{eq:transform}) preserves the canonical commutators:
\begin{align}
\left[\bar{\varphi}_{i}\left(x\right),\partial_{x'}\bar{\theta}_{k}\left(x'\right)\right]=i \pi\delta\left(x'-x\right)\delta_{ik}
\end{align}
(See App. \ref{sec:D_transform_properties}). The central mode $\bar{\theta}_0,\bar{\varphi}_0$ corresponds to the symmetric combination of the original fields
\begin{align}
\bar{\theta}_0=\sum_i \theta_i,\quad \bar{\varphi}_0=\frac{1}{2N+1}\sum_i \varphi_i.
\end{align}
We rescale it in a way that allows us to interpret $\bar{\theta}_0$ as the azimuthal angle of an effective spin operator. Namely, it encodes a $U(1)$ mode associated with a global rotation angle in the $XY$-plane. The remaining fields ($\bar{\theta}_i$ with $i\neq 0$) are rescaled in order for the most relevant terms to be of the form $\cos2\tilde{\theta}_i$:
\begin{align}
&\tilde{\theta}_{i}=\frac{1}{2}\bar{\theta}_{i},\quad\tilde{\varphi}_{i}=2\bar{\varphi}_{i}, \quad \forall i\neq0, \\
&\tilde{\theta}_{0}=\frac{1}{2N+1}\bar{\theta}_{0},\quad\tilde{\varphi}_{0}=\left(2N+1\right)\bar{\varphi}_{0}. \nonumber
\end{align}
In terms of the new fields, the Hamiltonian (\ref{eq:H_bosonized_long}) acquires the form
\begin{align}
\label{eq:H_new_basis_gen}
H&=H_{0}+H_{b}+H^{int} \nonumber \\
H_{b}&=\sum_{i\neq 0} H_b^i  \\
H^{int}&=H^{int}_{b,b}+H^{int}_{b,0}, \nonumber
\end{align}
where
\begin{widetext}
\begin{align}
\label{eq:H_new_basis}
H_{0}&=\frac{1}{2\pi}\int\mathrm{d}x\left[u_{0}K_{0}\left(\partial_{x}\tilde{\theta}_{0}\right)^{2}+\frac{u_0}{K_{0}} \left(\partial_{x}\tilde{\varphi}_{0}\right)^{2}\right]+\frac{1}{\pi}\tilde{B}_{0}\int\mathrm{d}x\partial_{x}\tilde{\varphi}_{0}\\
H_{b}^{i}&=\frac{1}{2\pi}\int\mathrm{d}x\left[u_{i}K_{i}\left(\partial_{x}\tilde{\theta}_{i}\right)^{2}+\frac{u_{i}}{K_{i}}\left(\partial_{x}\tilde{\varphi}_{i}\right)^{2}\right]+\frac{2g_{1}}{\left(2\pi\alpha\right)^{2}}\int\mathrm{d}x\cos 2\tilde{\theta}_{i}+\frac{1}{\pi}\tilde{B}_{i}\int\mathrm{d}x\partial_{x}\tilde{\varphi}_{i} \nonumber \\
H_{b,b}^{int}&=\sum_{i\neq j\neq0}\frac{1}{2\pi}\int\mathrm{d}x\left[u4K\partial_{x}\tilde{\theta}_{i}\left(DD^{T}\right)_{i,j}^{-1}\partial_{x}\tilde{\theta}_{j}+\partial_{x}\tilde{\varphi}_{i}\left\{ \frac{u}{4K}\left(DD^{T}\right)_{i,j}+\frac{J_{\perp}^{z}\alpha}{2\pi^{2}}\left(DCD^{T}\right)_{i,j}\right\} \partial_{x}\tilde{\varphi}_{j}\right]+\sum_{i\neq 0}\int\mathrm{d}x\frac{2g_{2}}{\left(2\pi\alpha\right)^{2}}\hat{O}_{i}^{c} \nonumber \\
H_{b,0}^{int}&=\frac{J_{\perp}^{z}\alpha}{2\pi^{2}}\frac{1}{2N+1}\int\mathrm{d}x\partial_{x}\tilde{\varphi}_{0}\left(\partial_{x}\tilde{\varphi}_{N}+\partial_{x}\tilde{\varphi}_{-N}\right)+\sum_{i\neq 0}\int\mathrm{d}x\left[-\frac{B_{i}}{\pi\alpha}\hat{O}_{i}^{a}+\frac{2g_{2}}{\left(2\pi\alpha\right)^{2}}\hat{O}_{i}^{b}+\frac{2g_{3}}{\left(2\pi\alpha\right)^{2}}\hat{O}_{i}^{d}\right]. \nonumber
\end{align}
\end{widetext}
Here $H_{0}$ is the Hamiltonian of the symmetric mode with
\begin{align}
u_{0}&=u\left(1+\frac{g_2}{u}\frac{N}{2N+1}\frac{2}{\pi}K\right)^{\frac{1}{2}}  \\
K_{0}&=\left(2N+1\right)K\left(\frac{1}{1+\frac{g_2}{u}\frac{N}{2N+1}\frac{2}{\pi}K}\right)^{\frac{1}{2}}, \nonumber
\end{align}
and $H_b$ describes a bath of 2N modes; $H^{i}_{b}$ is the Hamiltonian of the $i$'th mode, which has the form of a sine-Gordon model where the LL parameters of mode $i$ is
\begin{align}
\label{eq:u_and_K_i}
u_{i}&=u\left(\frac{2\left(N+\left|i\right|\right)\left(N-\left|i\right|+1\right)}{2N+1}\right)^{\frac{1}{2}}  \\
K_{i}&=K\left(\frac{8\left(N+\left|i\right|\right)\left(N-\left|i\right|+1\right)}{2N+1}\right)^{\frac{1}{2}} \nonumber.
\end{align}
$H_{b,b}^{int}$ and $H_{b,0}^{int}$ describe interactions within the bath modes and between the bath and the symmetric 0'th mode, respectively, in which
\begin{align}
\label{Oa}
&\hat{O}_{\pm i}^{a}=\cos\left(2\varphi_{\pm i}-2k_{F}x\right) \\
&=\begin{cases} \nonumber
\cos\left(\frac{2}{2N+1}\tilde{\varphi}_{0}+\tilde{\varphi}_{\pm i}-2k_{F}x\right) & ;i=N\\
\cos\left(\frac{2}{2N+1}\tilde{\varphi}_{0}+\tilde{\varphi}_{\pm i}-\tilde{\varphi}_{\pm\left(i+1\right)}-2k_{F}x\right) & ;1\leqslant i<N\\
\cos\left(\frac{2}{2N+1}\tilde{\varphi}_{0}-\tilde{\varphi}_{1}-\tilde{\varphi}_{-1}-2k_{F}x\right) & ;i=0
\end{cases}
\end{align}
\begin{align}
\label{Ob}
&\hat{O}_{\pm i}^{b}=\cos\left(2\varphi_{\pm i}+2\varphi_{\pm\left(i-1\right)}\right) \\
&=\begin{cases} \nonumber
\cos\left(\frac{4}{2N+1}\tilde{\varphi}_{0}+\tilde{\varphi}_{\pm\left(i-1\right)}\right) & ;i=N\\
\cos\left(\frac{4}{2N+1}\tilde{\varphi}_{0}+\tilde{\varphi}_{\pm\left(i-1\right)}-\tilde{\varphi}_{\pm\left(i+1\right)}\right) & ;1<i<N\\
\cos\left(\frac{4}{2N+1}\tilde{\varphi}_{0}-\tilde{\varphi}_{\mp i}-\tilde{\varphi}_{\pm\left(i+1\right)}\right) & ;i=1
\end{cases}
\end{align}
\begin{align}
\label{Oc}
&\hat{O}_{\pm i}^{c}=\cos\left(2\varphi_{\pm i}-2\varphi_{\pm\left(i-1\right)}\right) \\
&=\begin{cases} \nonumber
\cos\left(2\tilde{\varphi}_{\pm i}-\tilde{\varphi}_{\pm\left(i-1\right)}\right) & ;i=N\\
\cos\left(2\tilde{\varphi}_{\pm i}-\tilde{\varphi}_{\pm\left(i-1\right)}-\tilde{\varphi}_{\pm\left(i+1\right)}\right) & ;1<i<N\\
\cos\left(2\tilde{\varphi}_{\pm i}+\tilde{\varphi}_{\mp i}-\tilde{\varphi}_{\pm\left(i+1\right)}\right) & ;i=1
\end{cases}
\end{align}
\begin{align}
\label{Od}
&\hat{O}_{\pm i}^{d}=\cos\left(4\varphi_{\pm i}\right) \\
&=\begin{cases} \nonumber
\cos\left(\frac{4}{2N+1}\tilde{\varphi}_{0}+2\tilde{\varphi}_{\pm i}\right) & ;i=N\\
\cos\left(\frac{4}{2N+1}\tilde{\varphi}_{0}+2\tilde{\varphi}_{\pm i}-2\tilde{\varphi}_{\pm\left(i+1\right)}\right) & ;1\leqslant i<N\\
\cos\left(\frac{4}{2N+1}\tilde{\varphi}_{0}-2\tilde{\varphi}_{1}-2\tilde{\varphi}_{-1}\right) & ;i=0\; .
\end{cases}
\end{align}
Finally, all modes couple linearly to an effective magnetic field $\tilde{B}_i$ where
\begin{align}
\tilde{B}_{i}=\begin{cases}
\frac{1}{2N+1}\sum_{j}B_{j} & i=0\\
\frac{1}{2}\left(B_{i}-B_{i-sg\left(i\right)}\right) & i\neq0
\end{cases}\; .
\end{align}
This leads to a shift of the Fermi momenta $k_F\rightarrow k_F^i=k_F-k_B^i$, where
\begin{align}
\label{eq:effective_B}
2k_B^i\equiv -\left\langle \partial_{x}\tilde{\varphi}_{i}\right\rangle =\begin{cases}
\frac{\tilde{B}_0K_0}{u_0}\sim\frac{K}{u}\sum_{j} B_{j} & i=0\\
\frac{\tilde{B}_{i}K_{i}}{u_{i}}=\frac{\tilde{B}_{i}K}{u} & i\neq0
\end{cases}\; .
\end{align}

The Hamiltonian in the form Eqs. (\ref{eq:H_new_basis_gen}), (\ref{eq:H_new_basis}) is still rather complicated, as $H^{int}$ includes many coupling terms between the modes. However, since the cosine terms $\cos2\tilde{\theta}_i$ are highly relevant, the $\tilde{\theta}_i$-fields in the bath are ordered and their fluctuations are gapped. The masses of the fluctuation fields $\delta\tilde{\theta}_i$ are given to a good approximation by (see, e.g., App. E2 in Ref. [\onlinecite{Giamarchi}])
\begin{align}
m_{i}=\frac{1}{\alpha}\left(\frac{8g_{1}}{\pi u_{i}K_{i}}\right)^{\frac{1}{2-\frac{1}{K_{i}}}}.
\end{align} The marginal Gaussian terms of $H_{b,b}^{int}$ (which couple the various bath modes) are less relevant than the mass terms and can be treated perturbatively, leading to renormalization of the values of $m_i$. The low-energy dynamics is therefore dominated by the gapless symmetric (0) mode, and is described by $H_0$ with corrections resulting from its coupling to the gapped bath modes. In the next section we analyze these corrections systematically.

\section{Effective Theory}
\label{sec:EffectiveTheory}

Now, we are in a position to inspect the effect of the gapped modes on the symmetric $0$-mode, and derive an effective Hamiltonian $H_0^{eff}$ in terms of the fields $\tilde{\varphi}_0,\tilde{\theta}_0$. The bare Hamiltonian $H_0$ [Eq. (\ref{eq:H_new_basis})] describes a gapless mode, but couplings to the gapped modes may alter its behavior. To investigate it, we apply perturbation theory on $H^{int}_{b,0}$ up to second order in all coupling constants, and write the corresponding correction terms in the action. The resulting corrections $\delta S_0$ are then obtained using a mean-field approximation for the bath operators described by $H_b$ with renormalized parameters $u_i$, $K_i$ and $m_i$.

To first order, there are only two terms (the first term in $H_{b,0}^{int}$) that couple the symmetric mode to the $N$ and $-N$ gapped modes. They give rise to the following correction
\begin{align}
\label{eq:H_Gauss_int}
&\delta S_{int,1}^{Gauss}=-\frac{J_{\perp}^{z}\alpha}{\pi^{2}}\frac{1}{\left(2N+1\right)} \\
\times &\int\mathrm{d}\tau\int\mathrm{d}x\partial_{x}\tilde{\varphi}_{0}\left(\langle\partial_{x}\tilde{\varphi}_{N}\rangle+\langle\partial_{x}\tilde{\varphi}_{-N}\rangle\right), \nonumber
\end{align}
where the expectation values (evaluated with respect to $H_b^N$) are proportional to the "magnetic field" acting on the $N$, $-N$  modes [see Eq. (\ref{eq:effective_B})].
The resulting correction to $H_0$ can be absorbed in the definition of $\tilde{B}_0$ (the last term of $H_0$ Eq. (\ref{eq:H_new_basis})). Its main effect is to provide an additional shift of the effective Fermi wave vector $k_F^0$ by
\begin{equation}
\delta k_B^0\sim \left(\frac{J^z_\perp\alpha}{u}\right)\frac{K}{u}\left(B_N+B_{-N}\right)\; .
\end{equation}

We next turn to the second order corrections, which renormalize the Luttinger parameters of the symmetric mode. We start with the already mentioned first part of $H_{b,0}^{int}$. In addition to the effective magnetic field, it induces corrections to the LL parameters (see App. \ref{sec:LL_corrections}):
\begin{align}
\label{eq:S_Gauss_int_2nd_order}
&\delta S^{Gauss}_{int,2}\simeq \frac{1}{\left(2N+1\right)^2}\left(\frac{J_{\perp}^{z}\alpha}{\pi^{2}u_N}\right)^{2}\int\mathrm{d}^{2}r_1\int\mathrm{d}^{2}r_2 \nonumber \\
&\times\left\langle \partial_{x}\tilde{\varphi}_{\pm N}\left(r_1\right)\partial_{x}\tilde{\varphi}_{\pm N}\left(r_2\right)\right\rangle \partial_{x}\tilde{\varphi}_{0}\left(r_1\right)\partial_{x}\tilde{\varphi}_{0}\left(r_2\right),
\end{align}
where $\vec{r}_{1/2}=\vec{R}\pm\frac{1}{2}\vec{r}=\left(x_{1/2},u_N \tau_{1/2}\right)$. We perform integration over a relative coordinate $\vec{r}$, exploiting the fact that all the $\tilde{\varphi}_{i}$ fields correlations decay exponentially for gapped modes; this yields
\begin{align}
\label{eq:S_Gauss_int_final}
&\delta S^{Gauss}_{int,2}\simeq\frac{1}{\left(2N+1\right)^{2}}\left(\frac{g_{2}}{u_{N}}\right)^{2}\frac{\log\frac{1}{m_{N}\alpha}}{\pi^{3}} \nonumber \\
&\times\int\mathrm{d}x\int\mathrm{\mathrm{d}\tau}u_{N}\left(\partial_{x}\tilde{\varphi}_{0}\left(x,\tau\right)\right)^{2}.
\end{align}
Here and in the rest of the paper $m_{i}$ is the effective mass of the i'th bath mode, accounting for its renormalization  by the marginal coupling terms between the bath modes included in $H_{b,b}^{int}$ [Eq. (\ref{eq:H_new_basis})].

Other corrections to the LL parameters come from the second part of $H_{b,0}^{int}$, namely from the operators $\hat{O}_{i}^{a}$,  $\hat{O}_{i}^{b}$ and $\hat{O}_{i}^{d}$ [Eqs. (\ref{Oa}), (\ref{Ob}) and (\ref{Od})]. Details of the derivation are given in App \ref{sec:LL_corrections}. Let us start with $\hat{O}_{i}^{a}$, which results from coupling to the magnetic field:
\begin{align}
\label{eq:deltaS_N^a}
&\delta S_{\pm N}^{a,2} \simeq\left(\frac{B_{N}}{\pi\alpha u_N}\right)^{2}\int\mathrm{d}^{2}r_1\int\mathrm{d}^{2}r_2 \left\langle e^{i\tilde{\phi}_{\pm N}\left(r_1\right)}e^{-i\tilde{\phi}_{\pm N}\left(r_2\right)}\right\rangle \nonumber \\
&\times e^{i\frac{2}{2N+1}\tilde{\phi}_{0}\left(r_1\right)}e^{-i\frac{2}{2N+1}\tilde{\phi}_{0}\left(r_2\right)}e^{-2ik_F^N\left(x_1-x_2\right)}+h.c \nonumber \\
&\simeq-\left(\frac{B_{N}\alpha}{u_{N}}\right)^{2}\frac{1}{\left(2N+1\right)^{2}}\left(m_{N}\alpha\right)^{\frac{1}{2}}\frac{8}{\pi^{5}}  \\
&\times\int\mathrm{d}\tau\int\mathrm{d}xu_N\left[-3\left(\partial_{x}\tilde{\varphi}_{0}\left(x,\tau\right)\right)^{2}+\frac{1}{u_{N}^2}\left(\partial_{\tau}\tilde{\varphi}_{0}\left(x,\tau\right)\right)^{2}\right], \nonumber
\end{align}
and for $i=\pm 1\dots\pm \left(N-1\right)$ we get
\begin{align}
\label{eq:deltaS_i^a}
&\delta S_{\pm i}^{a,2} \simeq\left(\frac{\tilde{B}_{i}}{\pi\alpha u_i}\right)^{2}\int\mathrm{d}^{2}r_1\int\mathrm{d}^{2}r_2 \left\langle e^{i\tilde{\phi}_{\pm i}\left(r_1\right)}e^{-i\tilde{\phi}_{\pm i}\left(r_2\right)}\right\rangle \nonumber  \\
&\times \left\langle e^{i\tilde{\phi}_{\pm i+1}\left(r_1\right)}e^{-i\tilde{\phi}_{\pm i+1}\left(r_2\right)}\right\rangle \\
&\times e^{i\frac{2}{2N+1}\tilde{\phi}_{0}\left(r_1\right)}e^{-i\frac{2}{2N+1}\tilde{\phi}_{0}\left(r_2\right)}e^{-2ik_F^i\left(x_1-x_2\right)}+h.c \nonumber \\
&\simeq-\left(\frac{B_{i}\alpha}{u_{i}}\right)^{2}\frac{1}{\left(2N+1\right)^{2}}\left(m_{i}\alpha\right)\frac{2^{8}\log\frac{2k_{F}^{i}}{m_{i}}}{\pi^{5}} \nonumber \\
&\times\int\mathrm{d}\tau\int\mathrm{d}xu_i\left[3\left(\partial_{x}\tilde{\varphi}_{0}\left(x,\tau\right)\right)^{2}-\frac{1}{u_{i}^2}\left(\partial_{\tau}\tilde{\varphi}_{0}\left(x,\tau\right)\right)^{2}\right] \nonumber
\end{align}
where
\begin{align}
\label{eq:k_F^i}
2k_F^i\equiv 2(k_F-k_B^i)=\frac{\pi}{\alpha}-\frac{\tilde{B}_iK}{u}
\end{align}
[see Eq. (\ref{eq:effective_B})]. Here we have used a mean-field approximation where operators associated with different modes $i\neq 0$ are decoupled, and replaced by their expectation values. Also, $k_F^i$ is approximated by $k_F$ to keep the coefficients up to the second order in the coupling constants.   Corrections from  $\hat{O}_{i}^{b}$ have the same form as $\hat{O}_{i}^{a}$ apart form a pre-factor:
\begin{align}
\label{eq:deltaS_N^b}
&\delta S_{\pm N}^{b,2}\simeq-\left(\frac{g_{2}}{u_{N}}\right)^{2}\frac{1}{\left(2N+1\right)^{2}}\left(m_{N}\alpha\right)^{-\frac{7}{2}}\frac{2^{5}}{\pi^{3}}  \\
&\times\int\mathrm{d}\tau\int\mathrm{d}xu_i\left[\left(\partial_{x}\tilde{\varphi}_{0}\left(x,\tau\right)\right)^{2}+\frac{1}{u_{N}^2}\left(\partial_{\tau}\tilde{\varphi}_{0}\left(x,\tau\right)\right)^{2}\right], \nonumber
\end{align}
and for modes $i=\pm 1\dots\pm \left(N-1\right)$ we get
\begin{align}
\label{eq:deltaS_i^b}
&\delta S_{\pm i}^{b,2}\simeq-\left(\frac{g_{2}}{u_{i}}\right)^{2}\frac{1}{\left(2N+1\right)^{2}}\left(m_{i}\alpha\right)^{-3}\frac{2}{3\pi^{3}} \\
&\times\int\mathrm{d}\tau\int\mathrm{d}xu_i\left[\left(\partial_{x}\tilde{\varphi}_{0}\left(x,\tau\right)\right)^{2}+\frac{1}{u_{i}^2}\left(\partial_{\tau}\tilde{\varphi}_{0}\left(x,\tau\right)\right)^{2}\right]. \nonumber
\end{align}
Finally, we evaluate the correction $\delta S_{\pm i}^{d,2}$ resulting from the operators $\hat{O}_{i}^{d}$:
\begin{align}
\label{eq:deltaS_N^d}
&\delta S_{\pm N}^{d,2}\simeq-\left(\frac{g_{3}}{u_{N}}\right)^{2}\frac{1}{\left(2N+1\right)^{2}}\left(m_{N}\alpha\right)^{-2}\frac{4}{\pi^{4}} \\
&\times\int\mathrm{d}\tau\int\mathrm{d}xu_{N}\left[\left(\partial_{x}\tilde{\varphi}_{0}\left(x,\tau\right)\right)^{2}+\frac{1}{u_{N}^{2}}\left(\partial_{\tau}\tilde{\varphi}_{0}\left(x,\tau\right)\right)^{2}\right], \nonumber
\end{align}
and (for $i\neq N$)
\begin{align}
\label{eq:deltaS_i^d}
&\delta S_{\pm i}^{d,2}\simeq-\left(\frac{g_{3}}{u_{i}}\right)^{2}\frac{1}{\left(2N+1\right)^{2}}\left(m_{i}\alpha\right)^{-3}\frac{2^{5}c}{\pi^{3}}  \\
&\times\int\mathrm{d}x\int u_{i}\mathrm{d}\tau \left[\left(\partial_{x}\tilde{\varphi}_{0}\left(x,\tau\right)\right)^{2}+\frac{1}{u_{i}^{2}}\left(\partial_{\tau}\tilde{\varphi}_{0}\left(x,\tau\right)\right)^{2}\right], \nonumber
\end{align}
where $c$ is a real number of order unity.

Collecting the contributions from Eqs. (\ref{eq:S_Gauss_int_final})--(\ref{eq:deltaS_i^a}) and (\ref{eq:deltaS_N^b})--(\ref{eq:deltaS_i^d}), we obtain a correction ($\delta S^2_0$) to the bare Gaussian action ($S_0$, corresponding to $H_0$ from Eq. (\ref{eq:H_new_basis})), describing the symmetric mode. The resulting effective action can be cast in the form
\begin{align}
&S_0+\delta S_0^2 = \frac{1}{2\pi K_{0}^{\prime}} \\
&\int\mathrm{d}x\int u_{0}^{\prime}\mathrm{d}\tau\Bigg\{\left(\partial_{x}\tilde{\varphi}_{0}\left(x,\tau\right)\right)^{2}+\frac{1}{u_{0}^{\prime 2}}\left(\partial_{\tau}\tilde{\varphi}_{0}\left(x,\tau\right)\right)^{2}\Bigg\} \nonumber
\end{align}
with modified LL parameters  $K_{0}'$ and $u_{0}'$. The general expressions for the LL parameters are complicated (see \ref{eq:newLLpar1} and \ref{eq:newLLpar2} in App. \ref{sec:LL_corrections}). However, to leading order in the coupling constants, $K_{0}'$ can be cast in the compact form
\begin{align}
\label{eq:K_0^prime}
K_{0}^{\prime} &\simeq K\left(2N+1\right)\left(1-\delta\right), \\
\delta &
\equiv \sum_{\nu}\delta_{\nu}
\end{align}
where
\begin{align}
\label{deltas_def}
&\delta_{1}=\frac{g_{2}}{u}\frac{N}{2N+1}\frac{K}{\pi}\\
&\delta_{2}=4\pi K\frac{1}{2N+1}\sum_{i=1}^{N}\delta_{i}^{B}\left(3\frac{u_{i}}{u_{0}}-\frac{u_{0}}{u_{i}}\right) \nonumber \\
&\delta_{3}=-4\pi K\frac{1}{2N+1}\sum_{i=1}^{N}\delta_{i}^{g_{2}}\left(\frac{u_{i}}{u_{0}}+\frac{u_{0}}{u_{i}}\right) \nonumber \\
&\delta_{4}=-4\pi K\frac{1}{2N+1}\sum_{i=1}^{N}\delta_{i}^{g_{3}}\left(\frac{u_{i}}{u_{0}}+\frac{u_{0}}{u_{i}}\right) \nonumber
\end{align}
and
\begin{align}
\delta_{i}^{B}&\equiv\begin{cases}
\left(\frac{B_{N}\alpha}{u_{N}}\right)^{2}\left(m_{N}\alpha\right)^{\frac{1}{2}}\frac{2}{\pi^{5}}, & i=N\\
\left(\frac{B_{i}\alpha}{u_{i}}\right)^{2}\left(m_{i}\alpha\right)\frac{2^{6}}{\pi^{5}}\log\frac{\pi}{m_{i}\alpha} & i\neq N
\end{cases} \\
\delta_{i}^{g_{2}}&\equiv\begin{cases}
\left(\frac{g_{2}}{u_{N}}\right)^{2}2\left(m_{N}\alpha\right)^{-\frac{7}{2}}\frac{2}{\pi^{3}}, & i=N\\
\left(\frac{g_{2}}{u_{i}}\right)^{2}\left(m_{i}\alpha\right)^{-3}\frac{1}{6\pi^{3}} & i\neq N
\end{cases} \\
\delta_{i}^{g_{3}}&\equiv\begin{cases}
\left(\frac{g_{3}}{u_{N}}\right)^{2}\left(m_{N}\alpha\right)^{-2}\frac{4}{4\pi^{4}}, & i=N\\
\left(\frac{g_{3}}{u_{i}}\right)^{2}\left(m_{i}\alpha\right)^{-3}\frac{8}{\pi^{3}}c & i\neq N.
\end{cases}
\end{align}
The full expressions are given in App. \ref{sec:LL_corrections}, Eqs. (\ref{eq:newLLpar1}) and (\ref{eq:newLLpar2}). The sign of $\delta$ depends on the relative strength of the coupling constants.

Apart from modifying the LL parameters of the symmetric mode, coupling to the bath degrees of freedom may induce interactions. Recalling that all correlation functions of the disordered fields $\tilde{\varphi}_{i\neq0}$ decay exponentially, it is enough to examine only local terms of the form $\cos\left(\beta\tilde{\varphi}_{0}\right)$. To do so, we recall the following property of the expectation value:
\begin{align}
\label{eq:int_term_restrictions}
\sum_{j}A_{j}^{i}\neq0\Rightarrow\langle\prod_{i\neq0}e^{i\sum_{j}A_{j}^{i}\tilde{\varphi}_{i}\left(r_{j}\right)}\rangle=0.
\end{align}
This imposes many restrictions on possible interaction terms of the form
\begin{align}
\label{eq:general_int_term}
\left\langle \prod_{i}\left(\hat{O}_{i}^{a}\right)^{\left|n_{i}^{a}\right|}\left(\hat{O}_{i}^{b}\right)^{\left|n_{i}^{b}\right|}\left(\hat{O}_{i}^{c}\right)^{\left|n_{i}^{c}\right|}\left(\hat{O}_{i}^{d}\right)^{\left|n_{i}^{d}\right|}\right\rangle
\end{align}
where $n_{i}^{f}$ with $f=a,b,c,d$ count the number of operators of a specific type. Replacing all bath modes operators by their expectation values, this generates a term of the form $\cos\left(\beta\tilde{\varphi}_{0}\right)$ with $\beta=\frac{1}{2N+1}\sum_{i}\left(4n_{i}^{b}+2n_{i}^{a}+4n_{i}^{d}\right)$. After imposing all the restrictions [Eq. (\ref{eq:int_term_restrictions})], we obtain
\begin{align}
\label{eq:beta_definition}
\beta=2\times n,
\end{align}
where $n$ is an integer number and the full derivation is summarized in App. \ref{sec:Inter_terms}. The above analysis only states that $\cos\left(\beta\tilde{\varphi}_{0}\right)$ exists; the specific terms with $n=1,2$, which are the most relevant, are generated, e.g., from the following contributions
\begin{align}
\ensuremath{\left\langle\prod_{\nu=-N}^{N}\hat{O}_{\nu}^{a}}\right\rangle &\propto\cos\left(2\tilde{\varphi}_{0}+2k_{F}^0x\right) \nonumber \\
\left\langle\prod_{\nu=-N}^{N}\hat{O}_{\nu}^{d}\right\rangle &\propto\cos\left(4\tilde{\varphi}_{0}+4k_F^0 x\right).
\end{align}
Another property, that can be inferred from the analysis of the restrictions Eq. (\ref{eq:int_term_restrictions}), is that absolute anti-symmetry of the DW configuration, namely, the case of $B_i=-B_{-i}$, leads to the cancellation of the $\cos2\tilde{\varphi}_0$ terms. In the following analysis we assume a more generic case, where some asymmetry is always present.

The derivation described above leads to an effective action of the symmetric $0$-mode, of a relatively simple form:
\begin{align}
\label{eq:effectivModel}
S_{0}^{eff}&=S_{0}+\delta S_{0}^{2}\\
&+\frac{2g^{\prime}}{\left(2\pi\alpha\right)^{2}}\int\mathrm{d}x\int u_{0}^{\prime}\mathrm{d}\tau\cos\left(2\tilde{\varphi}_{0}+2k_{F}^{0}x\right) \nonumber \\
&+\frac{2g^{\prime\prime}}{\left(2\pi\alpha\right)^{2}}\int\mathrm{d}x\int u_{0}^{\prime}\mathrm{d}\tau\cos\left(4\tilde{\varphi}_{0}+4k_{F}^{0}x\right), \nonumber
\end{align}
where only one of two cosine terms will contribute depending on the corresponding oscillatory factors. Since $k_F^0=\pi/2\alpha-k_B^0$ where $k_B^0\propto \tilde{B}_0$ [see Eq. (\ref{eq:effective_B})], this depends crucially on the commensuration condition set by the effective magnetic field $\tilde{B}_0$. The exact expressions for the new coupling constants $g^{\prime}$ and $g^{\prime\prime}$ are complicated, however their values are of less significance to the low energy properties and are left unspecified.

\begin{figure}
\label{fig:PhaseDiagram1}
\includegraphics[scale=0.6]{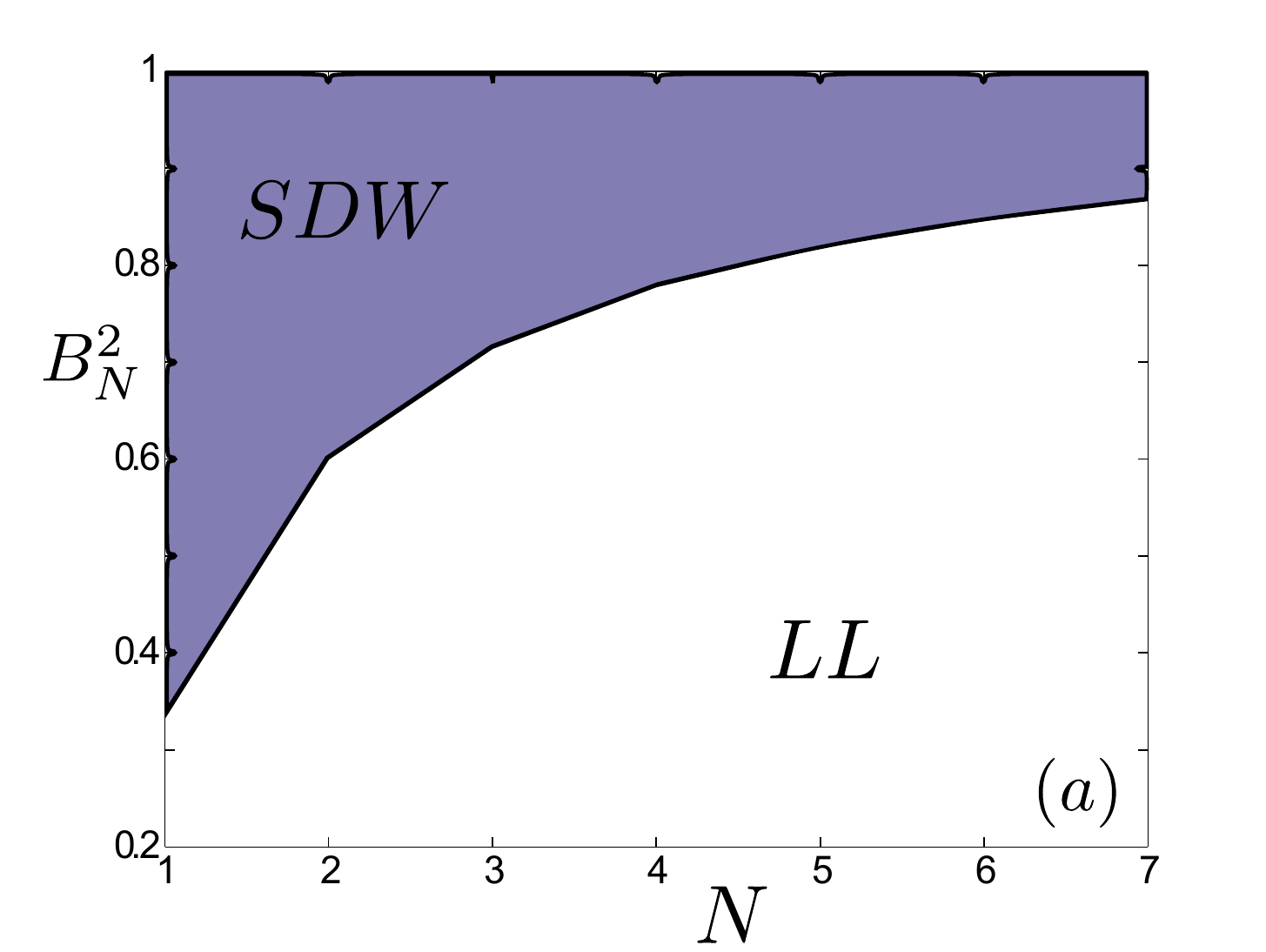}
\includegraphics[scale=0.6]{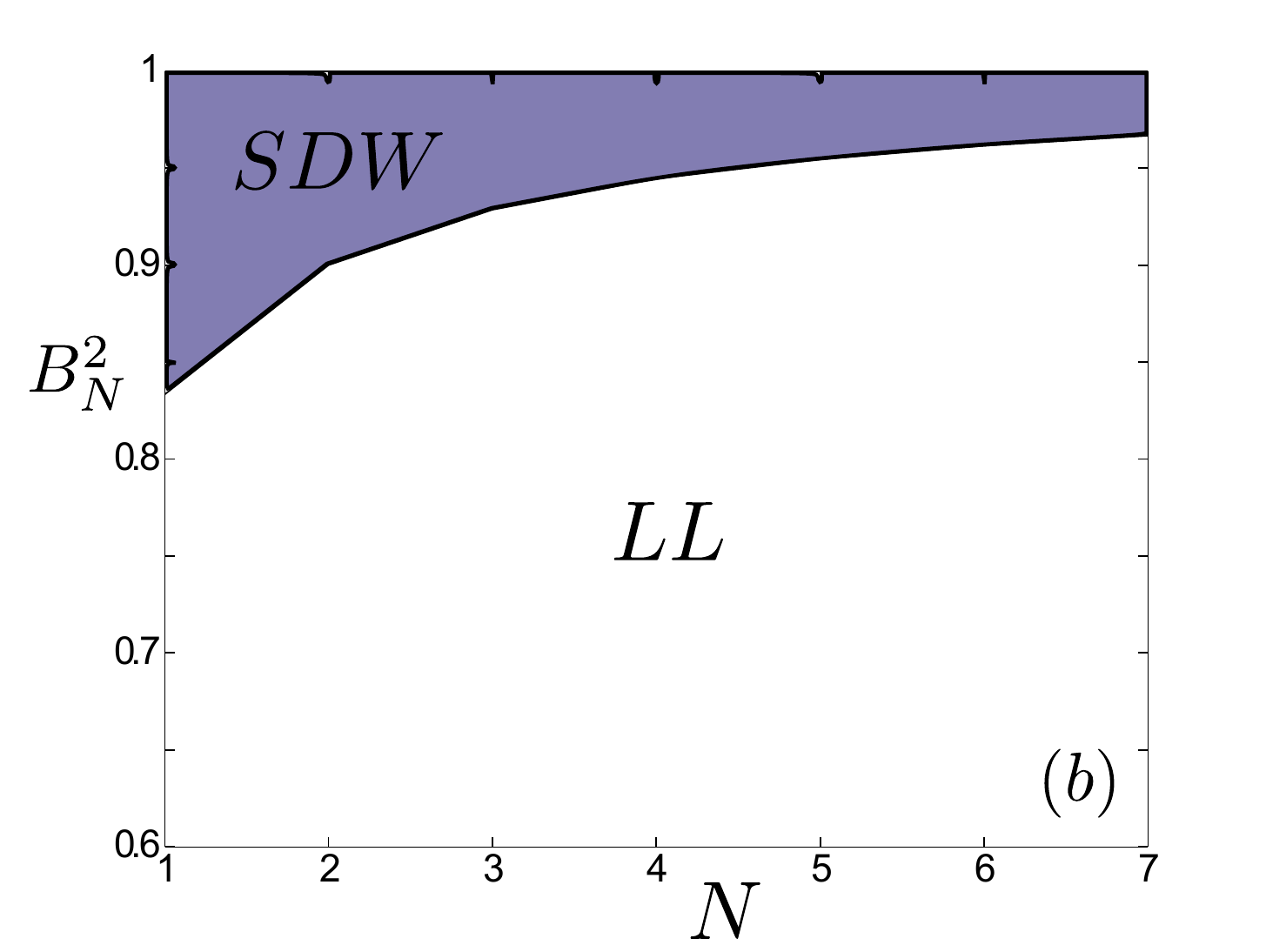}
\label{fig:PhaseDiagram2}
\caption{(Color online.) A schematic phase diagram. Here the magnetic field $B_N$ is given in units of $\frac{u_N}{\alpha}$. To plot the phase boundaries, we set $\delta_1=\delta_3=\delta_4=0$ and fix $K$ so that $4\pi K\frac{1}{2N+1}\sum_{i=1}^{N}\tilde{\delta}_{i}^{B}\left(3\frac{u_{i}}{u_{0}}-\frac{u_{0}}{u_{i}}\right) \sim 1$, where $\tilde{\delta}_{i}^{B}=\frac{u_N}{B_N \alpha}\delta^B_i$. The plot (a) is for the case of $\bar{B}\simeq\frac{u}{\alpha}\pi$ and (b) for the case of $\bar{B}\simeq 0$. }
\end{figure}

The dynamics of the effective theory Eq. (\ref{eq:effectivModel}) is determined by the relevance of cosine operators. Let us, first, inspect the $\cos\left(2\tilde{\varphi}_0+k_F^0x\right)$ term, which is relevant when
\begin{align}
\label{eq:cos2phi_condition}
K_0^{\prime}<2
\end{align}
provided the corresponding oscillatory factor is vanishing. The condition for that is
\begin{align}
2k_F^0\mod 4k_F \ll 2k_F,
\end{align}
leading to
\begin{align}
\label{eq:B_bar_asym}
\bar{B}\equiv\sum_j B_j \simeq\frac{u}{\alpha}\pi\sim J^{xy}_\parallel
\end{align}
[Eq. (\ref{eq:effective_B})]
where $\bar{B}$ may be seen as a measure of DW asymmetry. Plugging Eq. (\ref{eq:K_0^prime}) into the condition of Eq. (\ref{eq:cos2phi_condition}) and writing it in terms of $\delta$, we obtain the following relation
\begin{align}
\delta >\delta_{c2}= 1-\frac{2}{K\left(2N+1\right)}.
\end{align}
This inequality is satisfied only in the case where $\delta_2$ [Eq. (\ref{deltas_def})] is the larger correction to the parameter $K_0^{\prime}$; indeed, this corresponds to the case of strong Zeeman field $\bar{B}$ obeying Eq. (\ref{eq:B_bar_asym}). Under this condition, as $\delta$ is tuned above $\delta_{c2}$ one obtains a quantum phase transition of the Kosterlitz-Thouless type \cite{KosterlitzThouless} from a LL to a gapped phase where the $\cos\left(2\tilde{\varphi}_0\right)$ operator acquires a finite expectation value. Recalling the definition of the $S_z$ spin component in terms of Euler angles (see Eq. (\ref{eq:Spin_operators})) we can identify the order induced by the above mentioned cosine terms as spin-density wave (SDW) polarized along the $z$-axis. This result may be summarized in a schematic phase diagram presented in Fig. 2.

In the same fashion we derive a condition for the $\cos\left(4\tilde{\varphi}_{0}+4k_{F}^{0}x\right)$ term to be relevant:
\begin{align}
\delta >\delta_{c4}= 1-\frac{1}{2K\left(2N+1\right)}.
\end{align}
However this time the appropriate condition for cancelation of the oscillations
\begin{align}
\frac{\bar{B}}{u}\ll 2k_F
\end{align}
is easily satisfied for an almost symmetric DW, as well as for a random Zeeman field of zero average, even if its mean-square ($\frac{1}{2N+1}\sum_jB_j^2$) is rather large.
In these cases as well, the tuning of $\delta$ above $\delta_{c4}$ induces a transition to a SDW phase. The ordered spin structure in the two types of SDW phases are distinct: in the case of a strong Zeeman field with highly asymmetric profile ($\bar{B}\simeq\frac{u}{\alpha}\pi$), the period of SDW goes to infinity, which corresponds to ferromagnetic (FM) order along the ladder. On the other hand, in the case $\bar{B}\ll \frac{2k_F u}{\alpha}$ the period of the SDW is of order the lattice constant. Qualitatively, both phase diagrams Fig. 2(a) and Fig. 2(b) are the same but the range of the parameter space is considerably large in the case of FM order. Finally, in the case when non of the cosine terms is relevant, the low-energy dynamics  is determined solely by the LL parameters $K_0^{\prime}$, $u_0^{\prime}$ and one obtains a gapless LL dynamics.

To probe the SDW order explicitly, one may measure the local magnetization  $m(x)$ as a response to a local change $\delta B$ in the magnetic field at point $x$ along the DW:
\begin{align}
m=-\frac{\partial F}{\partial \left(\delta B\right)}\Bigg|_{\delta B =0}
\end{align}
where $F=-\frac{1}{\beta}\ln Z$ and $Z$ is a partition function calculated for the effective action Eq. (\ref{eq:effectivModel}). The exact dependence of the effective action on $\delta B$ is complicated, but the leading order comes from the induced modification of $g^{\prime}$ which acquires a correction linear in $\delta B$. This yields
\begin{align}
m(x) \propto\langle\cos\left(2\tilde{\varphi}_{0}+2k_{F}^{0}x\right)\rangle
\end{align}
which vanishes in the disordered LL phase. A transition to either of the SDW phases is marked by the emergence of a finite $m(x)$, which exhibits a modulation pattern along the DW on a scale dictated by the wave vector $k_F^0$. In particular, for
$\bar{B}\simeq \frac{u}{\alpha}\pi$, i.e. when $\cos\left(2\tilde{\varphi}_{0}\right)$ becomes relevant, the magnetization is constant in space, namely, FM order is established with total magnetization
\begin{align}
M \propto\int dx\langle\cos\left(2\tilde{\varphi}_{0}\right)\rangle\; .
\end{align}
On the other hand, for $\bar{B}\simeq 0$ where the order is dictated by the $\cos\left(4\tilde{\varphi}_{0}\right)$ term, the magnetization oscillates on a scale of $\frac{\pi}{2k_F^0}\sim \alpha$,
which is interpreted as AFM order.

\section{summary and concluding remarks}
\label{sec:sum_remarks}
In this paper we have studied a quasi-1D model for quantum spin fluctuations in a domain wall (DW) configuration generated by a non-uniform transverse field imposed on a ferromagnet with easy-plane $XXZ$-anisotropy. We find that the low-energy dynamics is dominated by a single soft mode propagating along the DW, reflecting the global $U(1)$ symmetry for rotating the spins in the $XY$-plane, which couples to a bath of gapped spin-fluctuation modes. An effective theory describing the quantum dynamics of this mode is derived, yielding in a large part of the parameters space a Luttinger liquid. The corresponding Luttinger parameter $K_0'$ grows systematically with the width of the DW, parametrized by the number of legs of the spin-ladder in our model. For an arbitrary finite width, a sufficiently strong mean-square value of the transverse Zeeman field can induce a quantum phase transition of the Kosterlitz-Thouless type \cite{KosterlitzThouless} into a SDW phase, where the spins are polarized along the local field direction and their fluctuations are gapped. However, with increasing width of the DW, the SDW phase progressively narrows. In the limit of infinite width (i.e. as the system becomes two-dimensional), one obtains a gapless LL dynamics where the effective Luttinger parameter $K_0'\rightarrow \infty$, restoring the Heisenberg FM limit. The gapless mode then becomes the Goldstone mode of the spontaneously symmetry-broken state.

In principle, a physical realization of our model is possible in a magnetic compound with anisotropic FM exchange interactions, in which case the transition into the SDW phase can be probed by measurement of magnetization or magnetic susceptibility. In addition, the same model applies for alternative realizations where the distinct phases are manifested by electric conduction properties. One prominent example is a low-dimensional superconducting (SC) device (e.g. a Josephson array) where the planar angle field $\theta_i$ represents the local phase of the complex order parameter, and $S_z$ the charge density operator so that the Zeeman field $B_z(X)$ corresponds to a gate voltage. In such systems, the transition to the SDW phase can be interpreted as a SC-insulator transition. Finally, in graphene devices and particularly bilayers in the quantum Hall regime, the conduction properties of polarized spin or isospin bulk phases are dominated by the 1D dynamic of DW's on the edges of the sample or the boundaries between domains of opposite polarization \cite{Kusum}. Due to the helical nature of these DW modes, the effective spin gap that opens in the SDW phase has quite the opposite     interpretation than in the SC analogue: it exponentially suppresses backscattering at low temperatures, leading to a nearly perfect 1D conductance \cite{us1416}.

{\it Acknowledgements -- } Useful discussions with H. A. Fertig, G. Murthy and R. Santos
are gratefully acknowledged. P. T. thanks the Bar-Ilan Institute for Nanotechnology and Advanced Materials for financial support during the academic year 2017. E. S. thanks the Aspen Center for Physics (NSF Grant No. 1066293) for its hospitality. This
work was supported by the US-Israel Binational Science Foundation
(BSF) grant 2012120, and the Israel Science Foundation (ISF)
grant 231/14.

\appendix

\section{Properties of the transformation $D$}
\label{sec:D_transform_properties}
In this Appendix we derive properties of the transformation $D$ defined in Eq.(\ref{eq:transform}). First, we show that it is indeed a canonical transformation
\begin{align}
&\left[\tilde{\varphi}_{i}\left(x\right),\partial_{x'}\tilde{\theta}_{k}\left(x'\right)\right]=\sum_{j,l}\left[\left(D^{T}\right)_{i,j}^{-1}\varphi_{j}\left(x\right),\partial_{x'}D_{k,l}\theta_{l}\left(x'\right)\right] \nonumber \\
&=\sum_{j,l}\left(D^{T}\right)_{i,j}^{-1}D_{k,l}\left[\varphi_{j}\left(x\right),\partial_{x'}\theta_{l}\left(x'\right)\right] \nonumber \\
&=\sum_{j,l}\left(D^{T}\right)_{i,j}^{-1}D_{k,l}\delta_{j,l}i\pi\delta\left(x'-x\right)\\
&=i\pi\delta\left(x'-x\right)\sum_{l}\left(D^{T}\right)_{i,l}^{-1}D_{k,l} \nonumber \\
&=i\pi\delta\left(x'-x\right)\sum_{l}\left(D\right)_{l,i}^{-1}D_{k,l}=i\pi\delta\left(x'-x\right)\delta_{i,k} \nonumber
\end{align}
For the treatment of the Hamiltonian (\ref{eq:H_Gauss}) we have chosen a specific from of the matrix $D$ (\ref{eq:transform}):
\begin{align}
D=\begin{pmatrix}\ddots & \vdots & \vdots & \vdots & \vdots & \vdots\\
\dots & 1 & -1 & 0 & 0 & 0 & \dots\\
\dots & 0 & 1 & -1 & 0 & 0 & \dots\\
\dots & 1 & 1 & 1 & 1 & 1 & \dots\\
\dots & 0 & 0 & -1 & 1 & 0 & \dots\\
\dots & 0 & 0 & 0 & -1 & 1 & \dots\\
 & \vdots & \vdots & \vdots & \vdots & \vdots & \ddots
\end{pmatrix}
\end{align}

The first part of the Hamiltonian $H_{Gauss}$ is changed by the symmetric matrix $DD^{T}$, as can be seen by exploring a general element of this matrix
\begin{align}
&\left(DD^{T}\right)_{a,b}=\sum_{c}\left(D\right)_{a,c}\left(D^{T}\right)_{c,b}\\
=&\sum_{c}\left(\delta_{a,c}+\delta_{a,0}-\delta_{a-sg\left(a\right),c}\right)\times\left(\delta_{b,c}+\delta_{b,0}-\delta_{b-sg\left(b\right),c}\right) \nonumber\\
=&\begin{cases}
\sum_{c}\left(\delta_{0,c}+1-\delta_{0,c}\right)\left(\delta_{0,c}+1-\delta_{0,c}\right) & ,a=b=0\\
\sum_{c}\left(\delta_{0,c}+1-\delta_{0,c}\right)\left(\delta_{b,c}+0-\delta_{b-sg\left(b\right),c}\right) & ,a=0,\forall b\neq0 \nonumber
\end{cases}\\
=&\begin{cases}
\sum_{c}1=2N+1 & ,a=b=0\\
\sum_{c}\delta_{b,c}-\sum_{c}\delta_{b-sg\left(b\right),c}=1-1=0 & ,a=0,\forall b\neq0 \nonumber
\end{cases}\; .
\end{align}
The Gaussian part of the interactions between modes, $H_{int}^{Gauss}$, is almost decoupled from the symmetric $0$-mode. To see this, we inspect terms of the matrix $DCD^{T}$:
\begin{align}
&\left(DCD^{T}\right)_{ab}=\sum_{i,j}D_{ai}C_{ij}\left(D^{T}\right)_{jb} \nonumber \\
&=\sum_{i,j}\left\{ \delta_{a,i}+\delta_{a,0}-\delta_{a-sg\left(a\right),i}\right\} \left\{ \delta_{i,j-1}+\delta_{i-1,j}\right\} \nonumber \\
&\times \left\{ \delta_{b,j}+\delta_{b,0}-\delta_{b-sg\left(b\right),j}\right\} \nonumber \\
&\left(DCD^{T}\right)_{00}=\sum_{i,j}\underbrace{\left\{ \delta_{i,j-1}+\delta_{i-1,j}\right\} }_{\text{{sum\,\ ovel\,\ all\,\ C\,\ elements}}}=2N \nonumber\\
&\left(DCD^{T}\right)_{0,N}=\sum_{i,j}\left\{ \delta_{0,i}+1-\delta_{0,i}\right\} \left\{ \delta_{i,j-1}+\delta_{i-1,j}\right\} \nonumber \\
&\times\left\{ \delta_{b,j}-\delta_{b-sg\left(b\right),j}\right\} \\
&=\sum_{i,j}\delta_{i,j-1}\left\{ \delta_{N,j}-\delta_{N-1,j}\right\} +\sum_{i,j}\delta_{i-1,j}\left\{ \delta_{N,j}-\delta_{N-1,j}\right\} \nonumber \\
&=\sum_{i}\underbrace{\left\{ \delta_{N,i+1}-\delta_{N-1,i+1}\right\} }_{1-1=0}+\sum_{i}\underbrace{\left\{ \delta_{N,i-1}-\delta_{N-1,i-1}\right\} }_{0-1}=-1 \nonumber\\
&\left(DCD^{T}\right)_{0,-N}=\sum_{i}\underbrace{\left\{ \delta_{-N,i+1}-\delta_{-N+1,i+1}\right\} }_{0-1} \nonumber \\
&+\sum_{i}\underbrace{\left\{ \delta_{-N,i-1}-\delta_{-N+1,i-1}\right\} }_{1-1=0}=-1 \nonumber\\
&\left(DCD^{T}\right)_{0,b\neq\pm N}=\sum_{i}\underbrace{\left\{ \delta_{b,i+1}-\delta_{b-sg\left(b\right),i+1}\right\} }_{1-1=0} \nonumber \\
&+\sum_{i}\underbrace{\left\{ \delta_{b,i-1}-\delta_{b-sg\left(b\right),i-1}\right\} }_{1-1=0}=0 \nonumber
\end{align}
and the resulting coupling to the symmetric mode is given by the first term in $H_{b,0}^{int}$ [Eq. (\ref{eq:H_new_basis})].

\section{Corrections to the LL parameters of the symmetric mode}
\label{sec:LL_corrections}
In this Appendix we present details for the derivation of corrections to the LL parameters. All the corrections come from the second order in perturbation theory. It is natural to treat this problem in the functional path integral language, namely, it is convenient to write the problem in terms of imaginary-time action. The induced corrections then result from re-exponentiation of a terminated perturbation series (i.e., a cumulant expansion). Symbolically, we can derive the perturbation theory in the following way:
\begin{align}
Z&=\int\mathrm{\prod_{i\neq0}D\tilde{\varphi}_{i}D\tilde{\varphi}_{0}}e^{-\left(S_{0}\left[\tilde{\varphi}_{0}\right]+S_{b}\left[\left\{ \tilde{\varphi}_{i\neq0}\right\} \right]+S_{int}\left[\left\{ \tilde{\varphi}_{i}\right\} \right]\right)} \nonumber\\
&=\int\mathrm{\prod_{i\neq0}D\tilde{\varphi}_{i}D\tilde{\varphi}_{0}}e^{-S_{0}\left[\tilde{\varphi}_{0}\right]}e^{-S_{b}\left[\left\{ \tilde{\varphi}_{i\neq0}\right\} \right]} \nonumber \\
&\times\left(1-S_{int}\left[\left\{ \tilde{\varphi}_{i}\right\} \right]+\frac{1}{2}S_{int}^{2}\left[\left\{ \tilde{\varphi}_{i}\right\} \right]+\dots\right)\\
&\simeq\int\mathrm{D\tilde{\varphi}_{0}}e^{-S_{0}\left[\tilde{\varphi}_{0}\right]}\left(1-\left\langle S_{int}\left[\left\{ \tilde{\varphi}_{i}\right\} \right]\right\rangle +\frac{1}{2}\left\langle S_{int}^{2}\left[\left\{ \tilde{\varphi}_{i}\right\} \right]\right\rangle \right) \nonumber\\
&=\int\mathrm{D\tilde{\varphi}_{0}}e^{-S_{0}\left[\tilde{\varphi}_{0}\right]-\left\langle S_{int}\left[\left\{ \tilde{\varphi}_{i}\right\} \right]\right\rangle +\frac{1}{2}\left\langle S_{int}^{2}\left[\left\{ \tilde{\varphi}_{i}\right\} \right]\right\rangle_{c} } \nonumber \\
&=\int\mathrm{D\tilde{\varphi}_{0}}e^{-S_{0}^{eff}\left[\tilde{\varphi}_{0}\right]} \nonumber
\end{align}
where $\left\langle S_{int}^{2}\left[\left\{ \tilde{\varphi}_{i}\right\} \right]\right\rangle_{c}=\left\langle S_{int}^{2}\left[\left\{ \tilde{\varphi}_{i}\right\} \right]\right\rangle-\left\langle S_{int}\left[\left\{ \tilde{\varphi}_{i}\right\} \right]\right\rangle^2$, and all expectation values are with respect to the bath action $S_{b}\left[\left\{ \tilde{\varphi}_{i\neq0}\right\} \right]$.

To evaluate the expectation value $\left\langle S_{int}^{2}\left[\left\{ \tilde{\varphi}_{i}\right\} \right]\right\rangle$, we need expressions for various correlation functions of the $\tilde{\varphi}_{i}$ fields which are strongly fluctuating when their dual fields $\tilde{\theta}_i$ are ordered. These are given by \cite{Gogolin}
\begin{align}
&\left\langle e^{i2\tilde{\varphi}_{i}\left(r_{1}\right)}e^{-i2\tilde{\varphi}_{i}\left(r_{2}\right)}\right\rangle \approx\left(m_i\alpha\right)^{2}\left(K_{0}^{2}\left(m_i r\right)+K_{1}^{2}\left(m_i r\right)\right) \label{eq:2phi_corfun}\\
&\left\langle \partial_{x}\tilde{\varphi}_i \left(r_{1}\right)\partial_{x}\tilde{\varphi}_i \left(r_{2}\right)\right\rangle_c\approx\frac{m_{i}^{2}}{2}\left\{ K_{1}^{2}\left(m_i r\right)-K_{0}^{2}\left(m_i r\right)\right\} \label{eq:gradphi_corfun} \\
&\left\langle e^{i\tilde{\varphi}_{i}\left(r_{1}\right)}e^{-i\tilde{\varphi}_i \left(r_{2}\right)}\right\rangle \propto\left(m_i\alpha\right)^{\frac{1}{2}}K_{0}\left(m_i r\right),\label{eq:1phi_corfun}
\end{align}
where $m_{i}$ are the masses, $r=\left|\vec{r}_1-\vec{r}_2\right|$ and $K_{i}\left(x\right)$ are Bessel functions of the second kind. The correlation functions \ref{eq:2phi_corfun} and \ref{eq:gradphi_corfun} are obtained by fermionization of SG Hamiltonian (which becomes exact for $K_i=1$), whereas \ref{eq:1phi_corfun} is calculated by mapping SG Hamiltonian to quantum Ising chain for $r\gg \frac{1}{m_{i}}$. All these correlation functions exponentially suppress contributions from $r> \frac{1}{m_{i}}$. Hence, inside the integrals Eqs. (\ref{eq:S_Gauss_int_2nd_order}), (\ref{eq:deltaS_N^a}), (\ref{eq:deltaS_i^a}) we can make the approximation
\begin{align}
\label{eq:gradient_app_phi_0}
&\tilde{\varphi}_0\left(\vec{r}_{1}\right)-\tilde{\varphi}_0\left(\vec{r}_{2}\right)=\tilde{\varphi}_0\left(\vec{R}+\frac{1}{2}\vec{r}\right)-
\tilde{\varphi}_0\left(\vec{R}-\frac{1}{2}\vec{r}\right) \nonumber \\
&\simeq\nabla\tilde{\varphi}_0\left(\vec{R}\right)\cdot\vec{r}
\end{align}
which allows us to integrated over the relative coordinate $\vec{r}$. Also whenever we have a product of two correlation functions with different velocities we approximate both velocities to be equal, because velocities of consecutive modes are almost equal.

Employing Eqs. (\ref{eq:2phi_corfun}) through (\ref{eq:gradient_app_phi_0}), we next derive the various second order corrections to $S_0^{eff}$.
First, we derive corrections that come form Eq. (\ref{eq:S_Gauss_int_2nd_order}):
\begin{align}
&\delta S_{int,2}^{Gauss}=\left\langle \left(S_{int}^{Gauss}\right)^{2}\right\rangle_{c} \\
&=\frac{1}{\left(2N+1\right)^2}\left(\frac{J_{\perp}^{z}\alpha}{\pi^{2}}\right)^{2}\frac{1}{u_{N}^{2}}\int\mathrm{d^{2}}r\int\mathrm{d^{2}}RF\left(r\right)\left(\partial_{x}\tilde{\varphi}_{0}\left(\vec{R}\right)\right)^{2} \nonumber \\
&=\frac{1}{\left(2N+1\right)^2}\left(\frac{J_{\perp}^{z}\alpha}{\pi^{2}}\right)^{2}\frac{F}{u_{N}^{2}}\int\mathrm{d}x\int\mathrm{\mathrm{d}\tau}u_{N}\left(\partial_{x}\tilde{\varphi}_{0}\left(x,\tau\right)\right)^{2} \nonumber
\end{align}
where, following the integration over the relative coordinate $\vec{r}$, we obtain 
\begin{align}
&F=\int\mathrm{d^{2}}rF\left(r\right)=\frac{m_N^{2}}{2}\int\mathrm{d^{2}r}\left(K_{1}^{2}\left(m_N r\right)-K_{0}^{2}\left(m_N r\right)\right) \nonumber \\
&\simeq2\pi\frac{1}{2}\left\{\ln \frac{1}{m_N\alpha}-\frac{1}{2}\right\}\simeq \pi \ln \frac{1}{m_N\alpha}
\end{align}
leading to Eq. (\ref{eq:S_Gauss_int_2nd_order}).
Similarly, we evaluate the contribution from the $O_{i}^{d}$ corrections, yielding
\begin{align}
&\delta S_{\pm N}^{d,2} =\left(\frac{2g_{3}}{\left(2\pi\alpha\right)^{2}}\right)^{2}\int\mathrm{d}^{2}r_1\int\mathrm{d}^{2}r_2 \left\langle e^{i2\tilde{\phi}_{\pm N}\left(r_1\right)}e^{-i2\tilde{\phi}_{\pm N}\left(r_2\right)}\right\rangle \nonumber \\
&\times e^{-2ik_{F}^{N}\left(x_{1}-x_{2}\right)} e^{i\frac{4}{2N+1}\tilde{\phi}_{0}\left(r_1\right)}e^{-i\frac{4}{2N+1}\tilde{\phi}_{0}\left(r_2\right)}+h.c \nonumber \\
&=-\left(\frac{2g_{3}}{\left(2\pi\alpha\right)^{2}}\right)^{2}\frac{1}{u_{N}^{2}}\left(\frac{4}{2N+1}\right)^{2}\left(m_{N}\alpha\right)^{2} \\
&\int\mathrm{d}\vec{R}\left\{\left(\partial_{R_{x}}\tilde{\varphi}_{0}\left(\vec{R}\right)\right)^{2}F_x+\left(\partial_{R_{\tau}}\tilde{\varphi}_{0}\left(\vec{R}\right)\right)^{2}F_{\tau}\right\} \nonumber
\end{align}
where
\begin{align}
F_x=&\int r\mathrm{d}r\mathrm{d}\theta\left[K_{1}^{2}\left(m_{N}r\right)+K_{0}^{2}\left(m_{N}r\right)\right]e^{-2ik_{F}^{N}r\cos\theta}\left(r\cos\theta\right)^{2} \nonumber \\
F_{\tau}=&\int r\mathrm{d}r\mathrm{d}\theta\left[K_{1}^{2}\left(m_{N}r\right)+K_{0}^{2}\left(m_{N}r\right)\right]e^{-2ik_{F}^{N}r\cos\theta}\left(r\sin\theta\right)^{2}.
\end{align}
Performing integration over $\vec{r}$ we obtain the result of  Eq. (\ref{eq:deltaS_N^d}).
\begin{align}
&\delta S_{\pm i}^{d,2} =\left(\frac{2g_{3}}{\left(2\pi\alpha\right)^{2}}\right)^{2}\int\mathrm{d}^{2}r_1\int\mathrm{d}^{2}r_2 \\
&\left\langle e^{i2\tilde{\phi}_{\pm i}\left(r_1\right)}e^{-i2\tilde{\phi}_{\pm i}\left(r_2\right)}\right\rangle\left\langle e^{i2\tilde{\phi}_{\pm i+1}\left(r_1\right)}e^{-i2\tilde{\phi}_{\pm i+1}\left(r_2\right)}\right\rangle \nonumber \\
&e^{i\frac{4}{2N+1}\tilde{\phi}_{0}\left(r_1\right)}e^{-i\frac{4}{2N+1}\tilde{\phi}_{0}\left(r_2\right)}+h.c\simeq\frac{4^2}{\left(2N+1\right)^2}\left(\frac{2g_{3}}{\left(2\pi\alpha\right)^{2}}\right)^{2}\frac{F_{4}}{u_{N}^{2}} \nonumber \\
&\times\left(m_{i}\alpha\right)^4\int\mathrm{d}\tau\int\mathrm{d}xu_{i}\left[\left(\partial_{x}\tilde{\varphi}_{0}\left(x,\tau\right)\right)^{2}+\frac{1}{u_{i}^2}\left(\partial_{\tau}\tilde{\varphi}_{0}\left(x,\tau\right)\right)^{2}\right], \nonumber
\end{align}
where
\begin{align}
&F_{4}=\int_{\alpha}^{\infty}\mathrm{d^{2}r}\left(K_{0}^{2}\left(m_{i}r\right)+K_{1}^{2}\left(m_{i}r\right)\right)^2\left|r_{x/\tau}\right|^{2} \nonumber \\
&\simeq\pi m_{i}^{-4}\left(0.075+2\times0.325+\pi\ln\left(\frac{1}{m_{i}\alpha}\right)\right)
\end{align}
leading to Eq. (\ref{eq:deltaS_N^d}).
All of the above corrections can be summarized as a set of equations of new LL parameters of the symmetric mode $\tilde{\varphi}_0$:
\begin{widetext}
\begin{align}
\label{eq:newLLpar1}
\frac{u_{0}^{\prime}}{2\pi K_{0}^{\prime}}&
\simeq\frac{u_{0}}{2\pi K_{0}}+\frac{1}{\left(2N+1\right)^{2}}\left(\frac{g_{2}}{u_{N}}\right)^{2}\frac{1}{\pi^{3}}\log\frac{1}{m_{N}\alpha}u_{N}\\
&+3\left(\frac{B_{N}\alpha}{u_{N}}\right)^{2}\left(\frac{2}{2N+1}\right)^{2}\left(m_{N}\alpha\right)^{\frac{1}{2}}\frac{2}{\pi^{5}}u_{N}+3\sum_{i}\left(\frac{B_{i}\alpha}{u_{i}}\right)^{2}\left(\frac{2}{2N+1}\right)^{2}\left(m_{i}\alpha\right)\frac{2^{6}}{\pi^{5}}\log\frac{\pi}{m_{i}\alpha}u_{i} \nonumber\\
&-\left(\frac{g_{2}}{u_{N}}\right)^{2}\left(\frac{2}{2N+1}\right)^{2}2\left(m_{N}\alpha\right)^{-\frac{7}{2}}\frac{2}{\pi^{3}}u_{N}-\sum_{i}\left(\frac{g_{2}}{u_{i}}\right)^{2}\left(\frac{2}{2N+1}\right)^{2}\left(m_{i}\alpha\right)^{-3}\frac{1}{6\pi^{3}}u_{i} \nonumber\\
&-\left(\frac{g_{3}}{u_{N}}\right)^{2}\left(\frac{4}{2N+1}\right)^{2}\left(m_{N}\alpha\right)^{-2}\frac{1}{4\pi^{4}}u_{N}-\sum_{i}\left(\frac{g_{3}}{u_{i}}\right)^{2}\left(\frac{4}{2N+1}\right)^{2}\frac{1}{\left(m_{i}\alpha\right)^{3}}\frac{2}{\pi^{3}}cu_{i} \nonumber
\end{align}
\end{widetext}
\begin{widetext}
\begin{align}
\label{eq:newLLpar2}
\frac{1}{2\pi K_{0}^{\prime}u_{0}^{\prime}}
&\simeq\frac{1}{2\pi K_{0}u_{0}}\\
&-\left(\frac{B_{N}\alpha}{u_{N}}\right)^{2}\left(\frac{2}{2N+1}\right)^{2}\left(m_{N}\alpha\right)^{\frac{1}{2}}\frac{2}{\pi^{5}}\frac{1}{u_{N}}-\sum_{i}\left(\frac{B_{i}\alpha}{u_{i}}\right)^{2}\left(\frac{2}{2N+1}\right)^{2}\left(m_{i}\alpha\right)\frac{2^{6}}{\pi^{5}}\log\frac{\pi}{m_{i}\alpha}\frac{1}{u_{i}} \nonumber \\
&-\left(\frac{2g_{2}}{u_{N}}\right)^{2}\left(\frac{2}{2N+1}\right)^{2}2\left(m_{N}\alpha\right)^{-\frac{7}{2}}\frac{2}{\pi^{3}}\frac{1}{u_{N}}-\sum_{i}\left(\frac{g_{2}}{u_{i}}\right)^{2}\left(\frac{2}{2N+1}\right)^{2}\left(m_{i}\alpha\right)^{-3}\frac{1}{6\pi^{3}}\frac{1}{u_{i}} \nonumber \\
&-\left(\frac{g_{3}}{u_{N}}\right)^{2}\left(\frac{4}{2N+1}\right)^{2}\left(m_{N}\alpha\right)^{-2}\frac{1}{4\pi^{4}}\frac{1}{u_{N}}-\sum_{i}\left(\frac{g_{3}}{u_{i}}\right)^{2}\left(\frac{4}{2N+1}\right)^{2}\frac{1}{\left(m_{i}\alpha\right)^{3}}\frac{2}{\pi^{3}}c\frac{1}{u_{i}} \nonumber
\end{align}\; ,
\end{widetext}
from which we deduce the final expressions for $K_{0}^{\prime}$, $u_{0}^{\prime}$.

\section{Derivation of induced interaction terms}
\label{sec:Inter_terms}
In this Appendix we summarize the treatment of the interaction terms in Eq. (\ref{eq:H_new_basis}), and analyze the restrictions mentioned in Eq. (\ref{eq:int_term_restrictions}). Consider a general term of the form Eq. (\ref{eq:general_int_term}).
The minimal requirement for non-zero correlation functions is $\sum_{\alpha}A_{\alpha}^{i}=0$ with $A_{\alpha}^{i}$ corresponding to $n_{i}^{f}$, leading to $2N$ equations:
\begin{widetext}
\begin{align}
\label{eq:restrictions}
\mu=N;\quad\varphi_{\pm\mu}:\quad&n_{\pm\mu}^{a}-n_{\pm\left(\mu-1\right)}^{a}&-n_{\pm\left(\mu-1\right)}^{b}&\quad+0&+2n_{\pm\mu}^{c}&-n_{\pm\left(\mu-1\right)}^{c}&-0\quad&+2n_{\pm\mu}^{d}-2n_{\pm\left(\mu-1\right)}^{d}&=0 \nonumber \\
1<\mu<N;\quad\varphi_{\pm\mu}:\quad&n_{\pm\mu}^{a}-n_{\pm\left(\mu-1\right)}^{a}&-n_{\pm\left(\mu-1\right)}^{b}&+n_{\pm\left(\mu+1\right)}^{b}&+2n_{\pm\mu}^{c}&-n_{\pm\left(\mu-1\right)}^{c}&-n_{\pm\left(\mu+1\right)}^{c}&+2n_{\pm\mu}^{d}-2n_{\pm\left(\mu-1\right)}^{d}&=0 \nonumber\\
\mu=1;\quad\varphi_{\pm\mu}:\quad&n_{\pm\mu}^{a}\quad-n_{0}^{a}&-n_{\pm\mu}^{b}\quad&+n_{\pm\left(\mu+1\right)}^{b}&+2n_{\pm\mu}^{c}&\quad+n_{\mp\mu}^{c}&-n_{\pm\left(\mu+1\right)}^{c}&+2n_{\pm1}^{d}-2n_{0}^{d}&=0.
\end{align}
\end{widetext}
By the end of the day, we want to calculate $\beta$. To do so we have to evaluate sums over $n_{i}^{f}$'s starting with $\sum_{i}n_{i}^{a}$. Eq. (\ref{eq:restrictions}) may be rearranged to give a general term $n_{\pm\mu}^{a}$
\begin{widetext}
\begin{align}
\label{eq:n_i^a_term}
&n_{\pm\mu}^{a}  \\
&=\begin{cases}
n_{0}^{a}+n_{\pm1}^{b}-n_{\pm2}^{b}-2n_{\pm1}^{c}+n_{\mp1}^{c}+n_{\pm2}^{c}-2n_{\pm1}^{d}+2n_{0}^{d} & \mu=1\\
\sum_{\alpha=1}^{\mu-1}n_{\pm\alpha}^{b}-\sum_{\alpha=3}^{\mu+1}n_{\pm\alpha}^{b}-2\sum_{\alpha=2}^{\mu}n_{\pm\alpha}^{c}+\sum_{\alpha=3}^{\mu+1}n_{\pm\alpha}^{c}+\sum_{\alpha=1}^{\mu-1}n_{\pm\alpha}^{c}-\sum_{\alpha=2}^{\mu}2n_{\pm\alpha}^{d}+\sum_{\alpha=1}^{\mu-1}2n_{\pm\alpha}^{d}+n_{\pm1}^{a} & 1<\mu<N\\
\sum_{\alpha=1}^{N-1}n_{\pm\alpha}^{b}-\sum_{\alpha=3}^{N}n_{\pm\alpha}^{b}-2\sum_{\alpha=2}^{N}n_{\pm\alpha}^{c}+\sum_{\alpha=3}^{N}n_{\pm\alpha}^{c}+\sum_{\alpha=1}^{N-1}n_{\pm\alpha}^{c}-\sum_{\alpha=2}^{N}2n_{\pm\alpha}^{d}+\sum_{\alpha=1}^{N-1}2n_{\pm\alpha}^{d}+n_{\pm1}^{a} & \mu=N
\end{cases} \; .\nonumber
\end{align}
Now we are in the position to perform the sum over $n_{\pm\mu}^{a}$ and write it terms of powers of $n_{\pm\mu}^{f}$ with $f\neq a$ and a few powers of $n_{\mu}^{a}$. The sum over $n_{\mu}^{a}$ simply gives
\begin{align}
\sum_{\nu}n_{\nu}^{a}&=Nn_{\pm1}^{a}+n_{0}^{a}+N\left(n_{\pm1}^{b}+n_{\pm2}^{b}\right)-2\sum_{\mu=1}^{N}n_{\pm\mu}^{b}+n_{\pm1}^{b}+\left(N-1\right)\left(n_{\pm1}^{c}-n_{\pm2}^{c}\right)-n_{\pm2}^{c}-\sum_{\mu=1}^{N}2n_{\pm\mu}^{d}+2n_{\pm1}^{d}N
\end{align}
and from that point it is straightforward to calculate $\beta$
\begin{align}
\frac{1}{2\left(2N+1\right)}\beta&=\sum_{\mu}\left(2n_{\mu}^{b}+n_{\mu}^{a}+2n_{\mu}^{d}\right)=\sum_{\mu=0,\pm1}^{\pm N}n_{\mu}^{a}+\sum_{\mu=0,\pm1}^{\pm N}\left(2n_{\mu}^{b}+2n_{\mu}^{d}\right)\\
&=N\left(2n_{0}^{a}+n_{\pm1}^{b}-n_{\pm2}^{b}-2n_{\pm1}^{c}+n_{\mp1}^{c}+n_{\pm2}^{c}-2n_{\pm1}^{d}+4n_{0}^{d}\right)+n_{0}^{a}+N\left(n_{\pm1}^{b}+n_{\pm2}^{b}\right)-2\sum_{\mu=1}^{N}n_{\pm\mu}^{b}+n_{\pm1}^{b} \nonumber\\
&+\left(N-1\right)\left(n_{\pm1}^{c}-n_{\pm2}^{c}\right)-n_{\pm2}^{c}-\sum_{\mu=1}^{N}2n_{\pm\mu}^{d}+2n_{\pm1}^{d}N+2\sum_{\mu=1}^{N}n_{\pm\mu}^{b}+2\sum_{\mu=0,\pm1}^{\pm N}n_{\mu}^{d} \nonumber\\
&=\left(2N+1\right)n_{0}^{a}+\left(2N+1\right)n_{\pm1}^{b}+\left(-3N+N-1\right)n_{\pm1}^{c}+\left(N-N+1-1\right)n_{\pm2}^{c}+4Nn_{0}^{d}+2n_{0}^{d} \nonumber\\
&=\left(2N+1\right)\left(n_{0}^{a}+n_{\pm1}^{b}-n_{\pm1}^{c}+2n_{0}^{d}\right). \nonumber
\end{align}
leading to the result Eq. (\ref{eq:beta_definition}).
\end{widetext}


\begin{thebibliography}{235}
\expandafter\ifx\csname natexlab\endcsname\relax\def\natexlab#1{#1}\fi
\expandafter\ifx\csname bibnamefont\endcsname\relax
  \def\bibnamefont#1{#1}\fi
\expandafter\ifx\csname bibfnamefont\endcsname\relax
  \def\bibfnamefont#1{#1}\fi
\expandafter\ifx\csname citenamefont\endcsname\relax
  \def\citenamefont#1{#1}\fi
\expandafter\ifx\csname url\endcsname\relax
  \def\url#1{\texttt{#1}}\fi
\expandafter\ifx\csname urlprefix\endcsname\relax\def\urlprefix{URL }\fi
\providecommand{\bibinfo}[2]{#2}
\providecommand{\eprint}[2][]{\url{#2}}

\bibitem{LadderReview} E. Dagotto, Rep. Prog. Rhys. {\bf 62}, 1525 (1999); E. Dagotto and N.M. Rice, Science {\bf 271}, 618 (1996).

\bibitem{HaldaneConjecture} Haldane, F. D. M., Phys. Rev. Lett., {\bf 50}, 1153 (1983); arXiv:1612.00076.

\bibitem{Schultz}
H. J. Schultz, Phys. Rev. B \textbf{34}, 6372 (1986).

\bibitem{Affleck}
I. Affleck, J. Phys. Cond. Matter {\bf 1}, 3047 (1991).

\bibitem{Shelton} D. G. Shelton, A. A. Nersesyan and A. M. Tsvelik,
Phys. Rev. B \textbf{53}, 8521 (1996).

\bibitem{balents-2010}
L. Balents, Nature {\bf 464}, 199 (2010).

\bibitem{AFMMaterials} K. Kopinga, A.M.C. Tinus and W.J.M. deJonge, Phys. Rev. B., {\bf 25}, 4685 (1982); {\bf 29}, 2868 (1984).

\bibitem{FMLadder1} T. Vekua, G. I. Japaridze and H.J. Mikesha, Phys. Rev. B., {\bf 67}, 064419 (2003).

\bibitem{FMLadder2} T. Vekua, G. I. Japaridze and H.J. Mikesha, Phys. Rev. B., {\bf 70}, 014425 (2004).

\bibitem{SOplusCoulomb} G. Jackeli and G. Khaliullin, Phys. Rev. Lett., {\bf 102}, 017205 (2009).


\bibitem{KitaevModel} A. Kitaev, Ann. Phys. (N.Y.), {\bf 321}, 2 (2006).

\bibitem{QHFM}
S. M. Girvin and A. H. MacDonald in \textit{Perspectives in Quantum Hall Effects}, S. Das Sarma and A. Pinczuk, eds. (John Wiley \& Sons, 1997);
D.H. Lee and C.L. Kane, Phys. Rev. Lett. {\bf 64}, 1313 (1990);
S.L. Sondhi, A.Karlhede, S.A. Kivelson and E. H. Rezayi, Phys. Rev. B {\bf47}, 16419 (1993); H.A. Fertig, L Brey, R. C\^{o}t$\acute{e}$, A.H. MacDonald, Phys. Rev. B {\bf50}, 11018 (1994);
K. Moon, H. Mori, K. Yang, S. M. Girvin, A. H. MacDonald, L.
Zheng, D. Yoshioka and S.-C. Zhang, Phys. Rev. B {\bf 51}, 5138 (1995).

\bibitem{Kharitonov_bulk}
M. Kharitonov, Phys. Rev. B {\bf 85}, 155439 (2012).

\bibitem{SupercondChains} E. W. Carlson, D. Orgad, S. A. Kivelson, V. J. Emery, Phys. Rev. B., {\bf 62}, 3422 (2000).

\bibitem{Maher2013}
P. Maher, C. R. Dean, A. F. Young, T. Taniguchi, K. Watanabe, K. L.
Shepard, J. Hone and P. Kim, Nature Physics {\bf 9}, 154 (2013).

\bibitem{Young2013}
A. F. Young, J. D. Sanchez-Yamagishi, B. Hunt, S. H. Choi, K. Watanabe, T. Taniguchi, R. C. Ashoori, P. Jarillo-Herrero, Nature {\bf 505}, 528 (2014).

\bibitem{ShibataTakagi} Junya Shibata and Shin Takagi, Phys. Rev. B., {\bf 62}, 5719 (2000).

\bibitem{FertigBrey} H. A. Fertig and L. Brey, Phys. Rev. Lett., {\bf 97}, 116805 (2006).

\bibitem{us1416}
G. Murthy, E. Shimshoni and H. A. Fertig, Phys. Rev. B {\bf 90}, 241410(R) (2014); P. Tikhonov, E. Shimshoni, H. A. Fertig and G. Murthy, Phys. Rev. B {\bf 93}, 115137 (2016).

\bibitem{Mazo14}
V. Mazo, C.-W. Huang, E. Shimshoni, S. T. Carr and H. A. Fertig, Phys. Rev. B {\bf 89}, 121411(R) (2014); V. Mazo, C.-W. Huang, E. Shimshoni, S. T. Carr and H. A. Fertig, Phys. Scr. {\bf T165}, 014019 (2015).

\bibitem{Kusum}
K. Dhochak, E. Shimshoni and E. Berg, ``Spontaneous layer polarization and conducting domain walls in the quantum Hall regime of bilayer graphene", Phys. Rev. B {\bf 91}, 165107 (2015).

\bibitem{Kharitonov_new}
M. Kharitonov, S. Juergens, B. Trauzettel, Phys. Rev. B {\bf 94}, 035146 (2016).

\bibitem{SpinBooks}
A. Auerbach, {\it Interacting Electrons and Quantum Magnetism} (Springer, 1994); A. Altland and B. Simons, {\it Condensed Matter Field Theory}, 2nd edition (Cambridge University Press, 2010).

\bibitem{MazoFertigShimshoni} V. Mazo, H. A. Fertig and E. Shimshoni, Phys. Rev. B., {\bf 86}, 125404 (2012).

\bibitem{SpinRot} Here we use freedom of the global spin rotation in the $XY$ plane; see, e.g., Ref. [\onlinecite{Giamarchi}].


\bibitem{LutherPeshelHaldane} A. Luther and J. P. Peschel, Phys. Rev. B, {\bf 12}, 3908 (1975); F. D. M. Haldane, Phys. Rev. Lett., {\bf 45}, 1358 (1980).

\bibitem{Giamarchi} T. Giamarchi, \textit{Quantum Physics in One Dimension},
(Oxford, New York, 2004).

\bibitem{KosterlitzThouless}
J. M. Kosterlitz and D. J. Thouless, J. Phys. C: Solid State {\bf
6}, 1181 (1973).

\bibitem{Gogolin} A. O. Gogolin, A. A. Nersesyan and A. M. Tsvelik,
\textit{Bosonization and Strongly Correlated Systems} (Cambridge University
Press, 1998).


\end{thebibliography}
\end{document}